\documentstyle[prl,aps,epsfig,multicol]{revtex}

\begin{document}
\title{Quantum transport properties of ultrathin silver nanowires}
\author{Jijun Zhao $^a$, Buia Calin $^a$, Jie Han $^b$, Jian Ping Lu $^a$$*$}
\address{$^a$ Department of Physics and Astronomy, University of North Carolina at Chapel Hill, Chapel Hill, NC 27599\\
$^b$ Eloret Corp., NASA Ames Research Center, MS 229-1, Moffett Field, CA 95051}
\maketitle

\begin{abstract}

The quantum transport properties of the ultrathin silver nanowires are investigated. For a perfect crystalline nanowire with four atoms per unit cell, three conduction channels are found, corresponding to three $s$ bands crossing the Fermi level. One conductance channel is disrupted by a single-atom defect, either adding or removing one atom. Quantum interference effect leads to oscillation of conductance versus the inter-defect distance. In the presence of multiple-atom defect, one conduction channel remains robust at Fermi level regardless the details of defect configuration. The histogram of conductance calculated for a finite nanowire (seven atoms per cross section) with a large number of random defect configurations agrees well with recent experiment.


\end{abstract}

\begin{multicols}{2}

In recent years, metallic nanowires are of great interest as building blocks for nanoelectronic devices. Since the dimension of the metallic nanowires is comparable to the electron Fermi wavelength, the conductance is quantized in units of $G_0=2e^2/h$ \cite{1,2,3}. The number of conductance channels is determined by the number of electronic bands crossing the Fermi level ($E_F$) and sensitively depends on the nanowire geometry. The effect of conductance quantization and the related high sensitivity lead to potential applications such as single-atom switch \cite{4}, conductor \cite{5,6}, and chemical sensor \cite{7}. 

Most earlier works on the quantum conductance of metallic nanowires are based on the point contacts formed between metal electrodes \cite{1}. With mechanically controllable break junctions \cite{8,9} or tip-surface contact \cite{10,11,12,13} techniques, the statistical histograms of the conductance value for large number of contacts have been  recorded. However, the metallic nanowires obtained from nanoscale contact are limited by the short length and structural instability for practical applications. Other fabricating methods, such as reduction of metal compounds \cite{14}, ions irradiation \cite{15}, carbon nanotubes capillary growth \cite{16,17,18}, and template-aid synthesis \cite{19,20} have been introduced to generate much longer nanowires with well-defined structures. Understanding the transport properties of these long and nearly freestanding metallic nanowires are important for their future applications in nanoelectronics. Recently, Kim's group has successfully obtained ultrathin single-crystalline silver nanowire arrays \cite{20}. The silver nanowires with 0.4 nm width (only four atoms on the cross section) and  ${\mu}$m-scale length are grown inside the pores of organic templates. In this letter, we investigate the transport properties of the ultrathin silver nanowires and discussed the effect of defects on the conductances. 

The geometry optimization for the silver nanowires are performed by using {\em ab initio} plane-wave ultrasoft pseudopotential method \cite{21}. Both atomic positions and cell parameters for the nanowires (perfect crystalline or with structural defects) were fully relaxed at level of local density approximation (LDA). We adopt the initial configurations proposed in Ref.[20], that is, fcc-like crystalline structure with wire axis along the [110] crystal direction. The equilibrium structures of a perfect silver nanowire are illustrated in Fig.1. The calculated atomic layer spacing along the wire axis (2.75 {\AA}) agree well with the experimental result \cite{20}. In this one-dimensional periodic structure, the four atoms in the unit cell belong to two inequivalent sites ($h$ and $l$). Two atoms have lower coordination number ($l$: $Z=6$); the other two atoms have higher coordination ($h$: $Z=7$) (see Fig.1(b)). 

\begin{figure}
\centerline{
\epsfxsize=2.4in \epsfbox{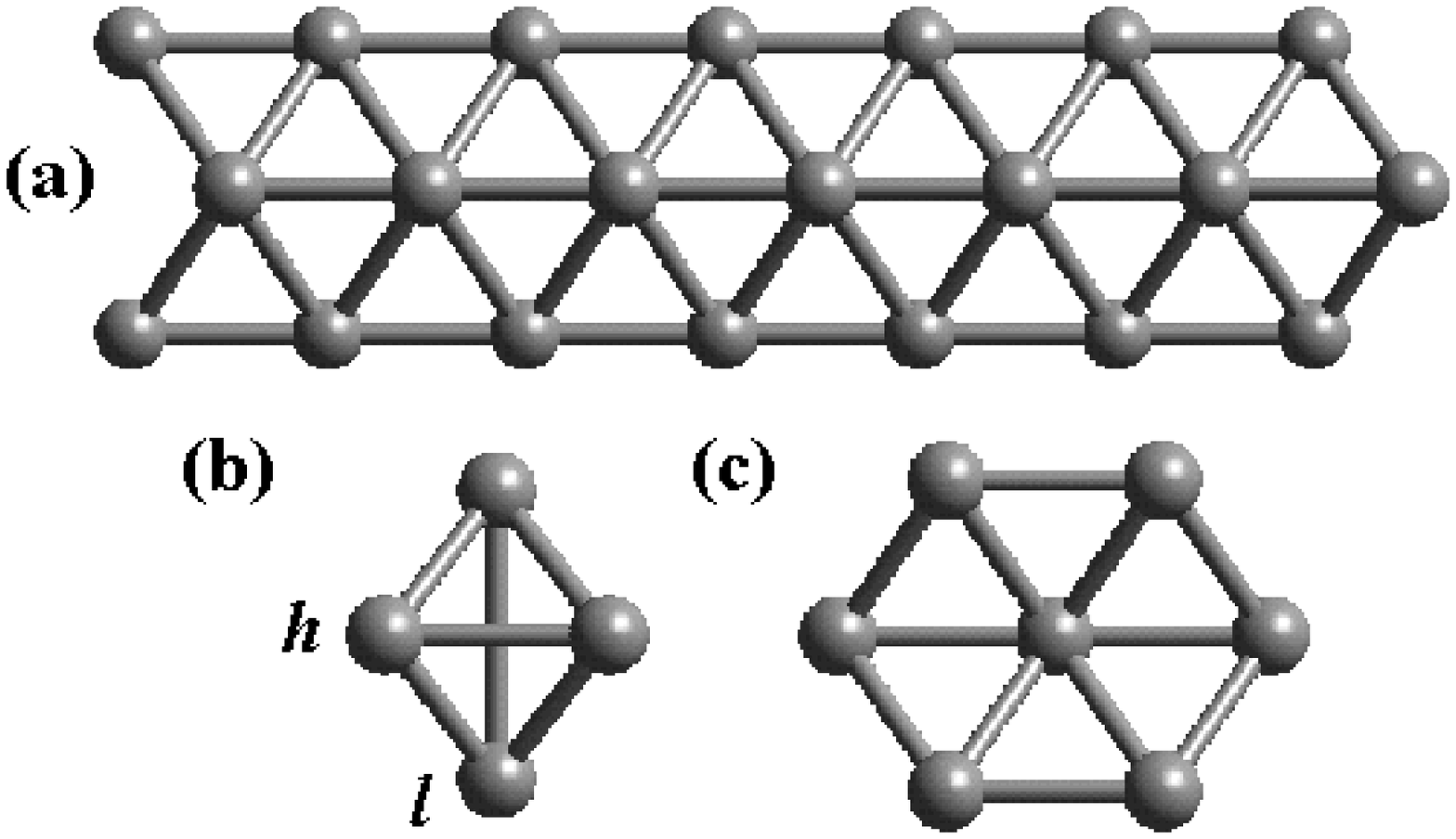}
}
\caption{Structures of crystalline silver nanowires with wire axis along [110] fcc  direction: (a), (b): side and top views of a thinner wire with four atoms per unit cell (as observed in Ref.[20]). The two inequivalent sites ($h$, $l$) with lower or higher coordination are labeled in (b). (a), (c): side and top views of a thicker wire with seven atoms per unit cell, which was suggested in Ref.[31].}
\end{figure}
\vspace{-0.1in}

It is known that the $4d$ orbitals in Ag atom act almost like innershell core orbitals \cite{22}. Our {\em ab initio} LDA band structure calculations also show that the $d$ bands are low-lying and have little influence on the $s$ electron bands around Fermi level. Therefore, the conduction electrons in silver should be reasonably described by a $s$-orbital tight-binding (TB) model, which had been used to describe the silver clusters up to Ag$_{68}$ \cite{23}. The $s$-orbital TB Hamiltonian is written as
$$
H=\sum_{i} \epsilon_i + \sum_{<ij>} t_{ij} 
$$
where the on-site energy $\epsilon_i= \epsilon_0+\alpha Z_i$ depends on the site coordination number $Z_i$. The hopping integrals $t_{ij}$ depend on the interatomic distance $r_{ij}$ between site $i$ and $j$, $t_{ij}=t_0(d_0/r_{ij})^2$ where $d_0$ is the  bulk nearest neighboring distance (2.89 {\AA}), $t_0$ is the $s$-orbital hopping parameter at the bulk distance $d_0$. It was found that parameter set $t_0$=0.89 eV and $\alpha$=0.26 eV can describe the silver clusters well \cite{23}. The same parameters are used in this work. In Fig.2, the electronic band structures of perfect silver wires near Fermi level by the $s$-orbital TB model are compared with those from accurate LDA calculations. Fig.2 clearly shows that the TB model can describe the electronic structures of silver wire within 1 eV of $E_F$ reasonably well.

\begin{figure}
\vspace{0.7in}
\centerline{
\epsfxsize=3.0in \epsfbox{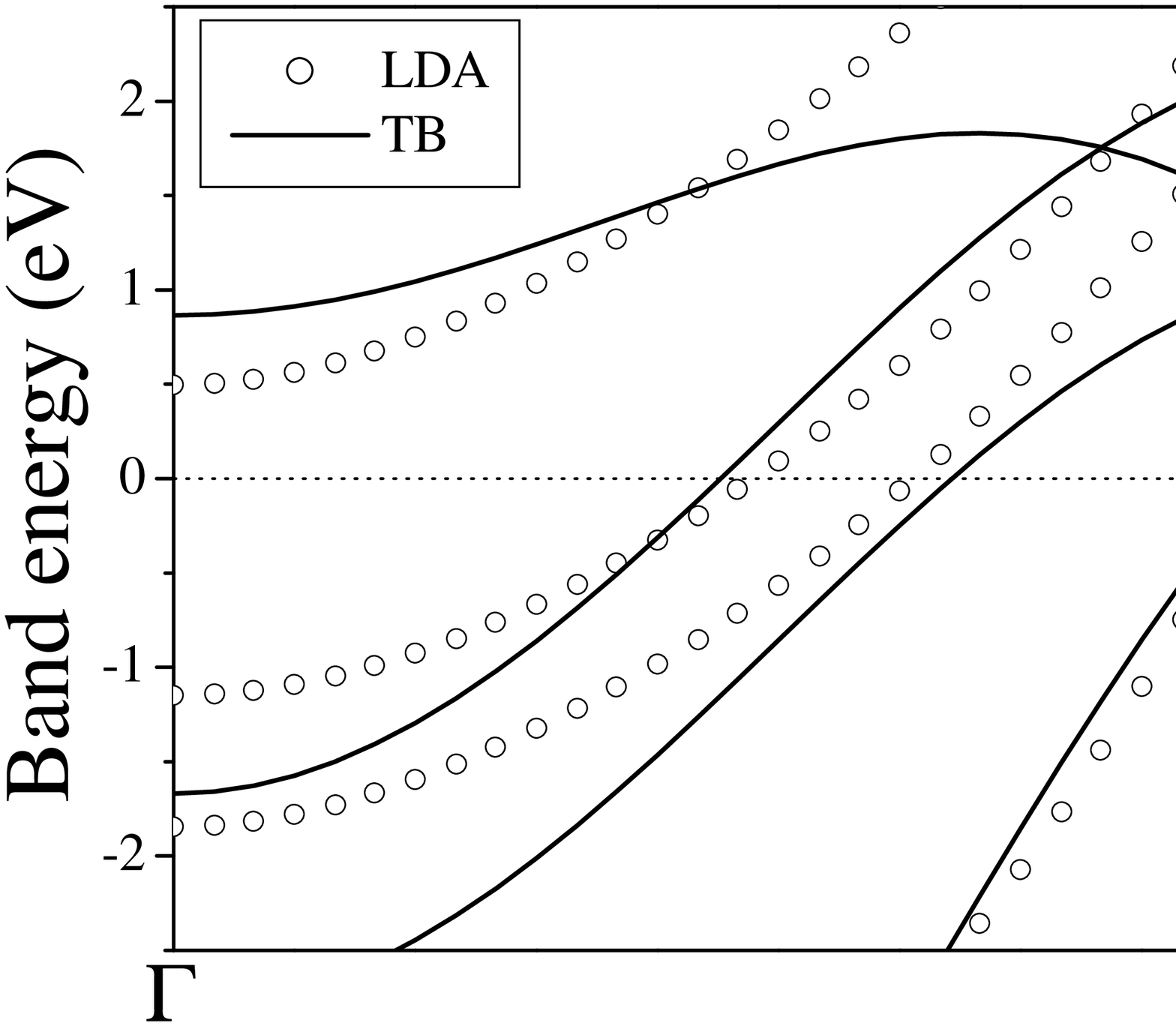}
}
\vspace{-0.65in}
\caption{Band structures of silver nanowires (Fig.1b) along the 1-D wire axis direction ($\Gamma$-$X$). The $s$-orbital tight-binding (solid lines) calculations reproduce most of the features from LDA (open circles). }
\end{figure}

We use the TB Hamiltonian to study the transport properties of silver nanowires via the surface Green's function matching method \cite{24,25}. The conductance $G$ is given by Landauer-B\"{u}ttiker formula with the transmission coefficient $T$ calculated from the surface Green's function. To mimic the nanowire of $\mu$m-scale length, the length of central nanowire section is chosen to be sufficient long such that the results are independent of the choice of length. The perfect crystalline nanowires of the same size were used as leads in the two sides. Similar tight-binding based surface Green's function methods have been successfully applied to the transport properties of metallic nanowires and carbon nanotubes \cite{26,27,28,29}. 

The calculated conductance of the perfect crystalline silver nanowire with four atoms per unit cell is shown in Fig.3. Three conduction channels at the Fermi level are found, corresponding to the three $s$ electronic bands crossing the Fermi level as predicted by both TB and LDA calculations (see Fig.2). As the ultrathin silver nanowire has only few atoms on the cross section, it is inevitable that there will be some structural defects on the wire. Two type of defects, vacancy or adatom are considered in present work. In the case of vacancy, the missing atom can be from either the low-coordination ($l$) sites or high-coordination ($h$) sites (see Fig.1). Our results (Fig.3) clearly show that one conduction channel is disrupted by a single-atom defect (either vacancy or adatom), independent of the details of defect configurations. This effect can be understood by the scattering of conduction electrons by the defect-induced potential. Similar effect was found in the carbon nanotube \cite{29}. 

\begin{figure}
\vspace{0.5in}
\centerline{
\epsfxsize=3.0in \epsfbox{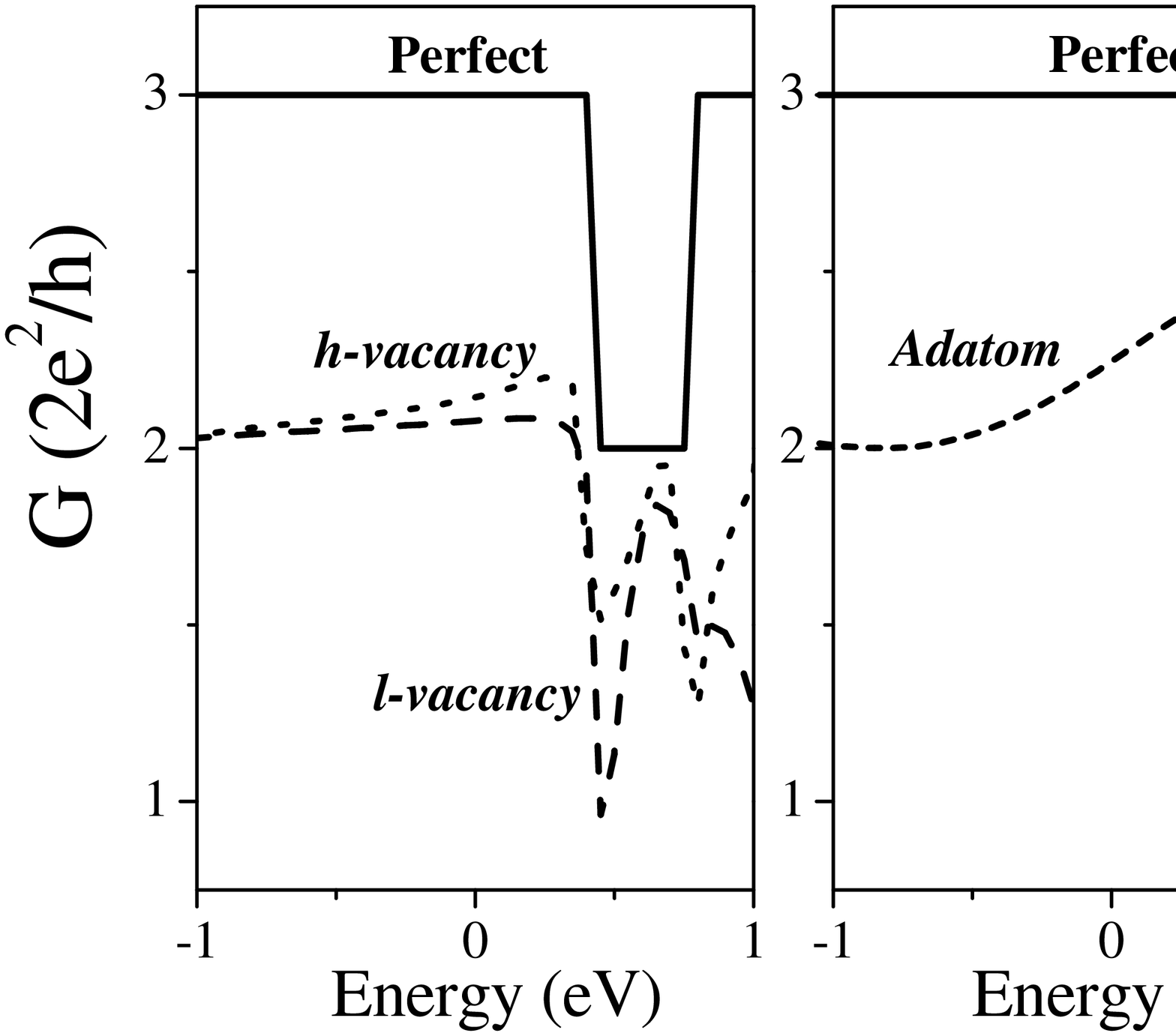}
}
\vspace{-0.7in}
\caption{Conductance of silver nanowires (Fig.1b) with single-atom defect. Solid line: perfect nanowires; dotted line (left): single-atom vacancy at high-coordination site; dashed line (left): single-atom vacancy at low-coordination site; short dashed line (right): adatom single-atom defect. In all the cases, one conduction channel is disrupted by a single-atom defect}
\end{figure}

\begin{figure}
\vspace{-0.1in}
\centerline{
\epsfxsize=3.0in \epsfbox{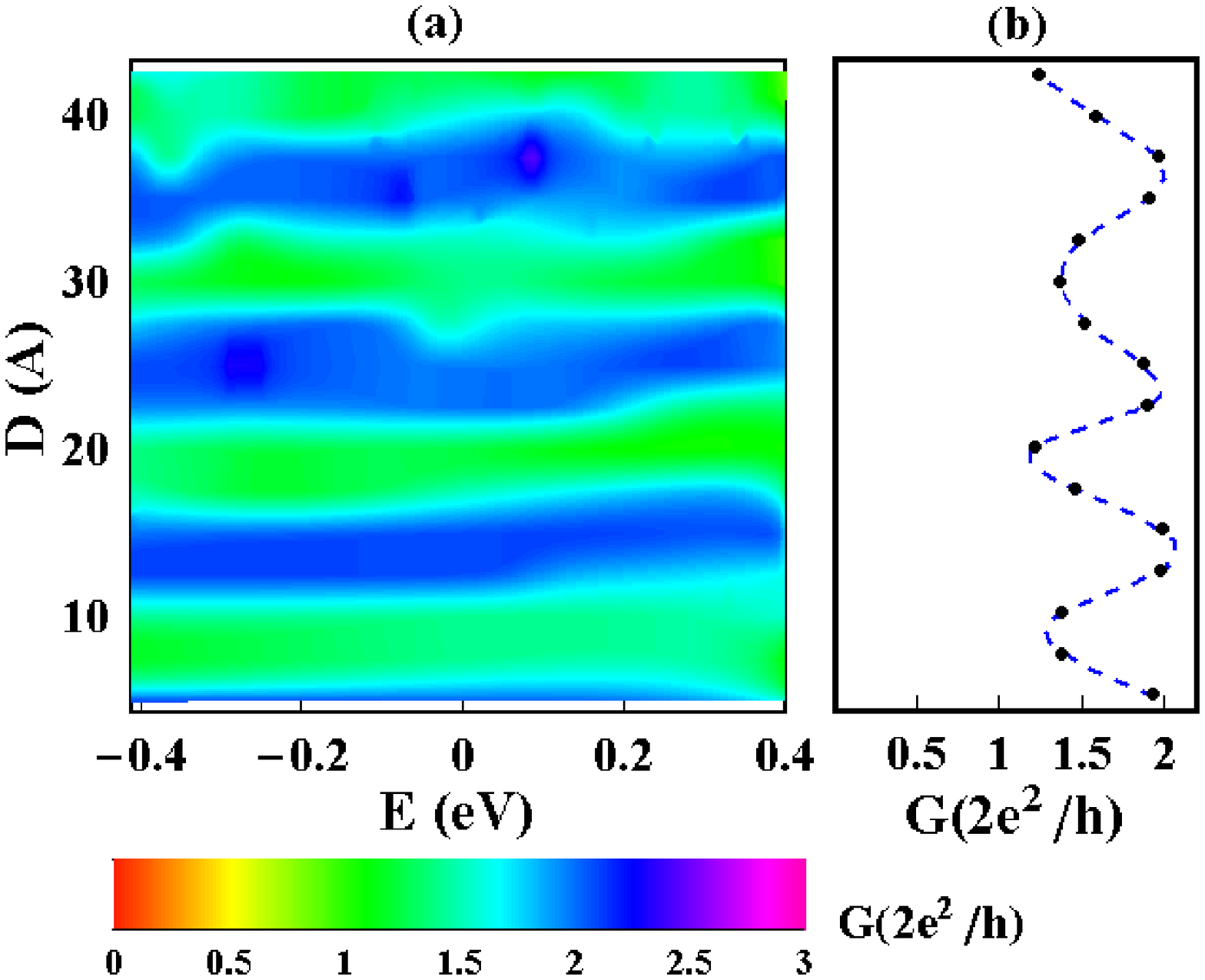}
}
\vspace{0.05in}
\caption{(a): 2-D contour plot of the conductance of silver nanowires (Fig.1b) vs. inter-$h$-vacancy distance and energy. (b) A cut of (a) through Fermi energy. Oscillation of conductance with the defect distance can be seen. }
\end{figure}

In addition to the effect of individual defect, we have also considered the situation of two separated single-atom defects. Oscillation of conductance with the defect distance is observed. An example as shown in Fig.4 is the 2-D contour plot of conductance versus defect distance and energy. The oscillation behavior can be understood by quantum interference of the conduction electronic states with oscillation scale related to Fermi wavelength. From our {\em ab initio} calculations, the density of conduction electrons in the silver wire shows similar oscillation with the same periodicity. Such quantum interference effects were also found in the carbon nanotubes \cite{28,30}.

\begin{figure}
\vspace{0.7in}
\centerline{
\epsfxsize=3.0in \epsfbox{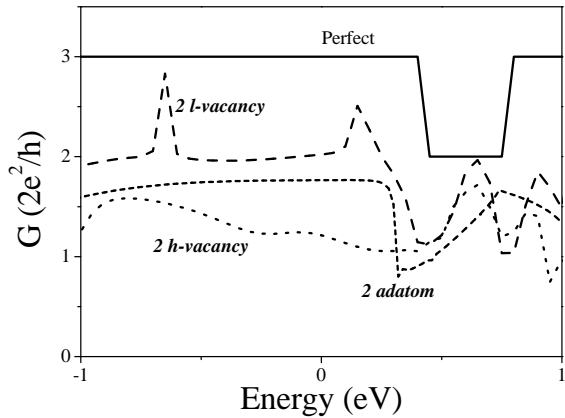}
}
\vspace{-0.65in}
\caption{Conductance of silver nanowires (Fig.1b) with two-atoms defect. Solid line: perfect nanowire; dotted line: defect with two-atoms vacancy at high-coordination site; dashed line: defect with two-atoms vacancy at low-coordination site; short dashed line: defect with two adatoms. One to two conduction channels are disrupted by the two-atoms defect at Fermi level.}
\end{figure}

We further investigate the other defect configurations such as multiple-atoms vacancies. In the cases of two-atoms vacancy (either both on the $h$ sites or both on the $l$ sites), substantial difference is found between different defect configurations (see Fig.5). At Fermi level, one to two conduction channels are disrupted by the two-atoms vacancy defect, sensitively depending on the local coordination environment of the defect site. It is reasonable to find that removing two atoms on the high-coordination site disrupts two conduction channels, while the absence of two atoms on the low-coordination site only reduce one conduction channel at Fermi energy. For individual defect with two adatoms, the result is between the case of two $h$ and two $l$ vacancy. Further removing one more atom will create a three-atoms vacancy with one atom link. Our calculations show that there is still one conduction channel open. In the earlier experimental studies on metallic contacts, it was usually found that the nanowire retain one conduction channel before breaking \cite{1}. Our results indicate that at least one conduction channel in the silver nanowire is rather robust. Hence, the ultrathin silver wire might serve as good interconnctor in nanoelectronics even with some structural defects formed during the fabrication process.

So far, there is no direct experimental measurement on the conductance of very long ultrathin silver nanowire. Our current studies can be related to a recent experiment on the quantum conductance of short silver nanowires generate by mechanical elongation \cite{31}. Similar to that in Ref.[20], rodlike nanowires along the [110] direction were observed. The global histogram recorded for the conductance of silver nanowires shows large peaks at 1 $G_0$, $\sim$2.4 $G_0$, and $\sim$4 $G_0$ \cite{31}. One possible atomic structure for the nanowire in their experiment is shown in Fig.1(c) with seven atoms per cross section, which can be built from the thinner wire in Fig.1(b). Using the same computational scheme, we have calculated the quantum conductance of such finite nanowire with seven unit cells in the central section ($\sim$ 2nm, comparable to experimental length). Without any defect, the nanowire has five conduction channels. The conductance drops to 4 $G_0$ with one single-atom defect, similar to that in the case of 4-atoms cross section nanowire discussed above. Our LDA calculations show four robust $s$ bands crossing the Fermi level, in consistent with TB results.

\begin{figure}
\vspace{0.5in}
\centerline{
\epsfxsize=3.0in \epsfbox{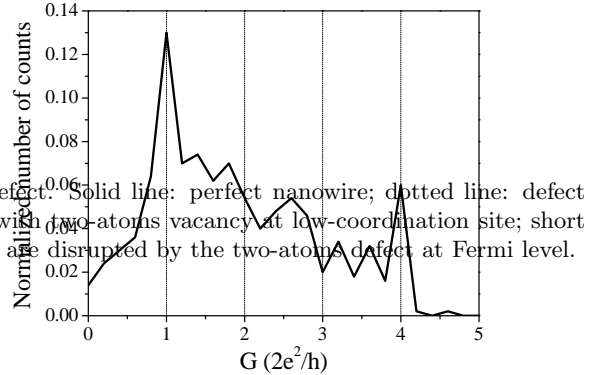}
}
\vspace{-0.7in}
\caption{Histogram of conductance of finite silver nanowire in Fig.1(c) with 500 random defect configurations. Three major peaks at 1 $G_0$, 2.6 $G_0$, and 4 $G_0$ are found}
\end{figure}

To compare with experimental conductance histogram, we generate a large number ($\sim$ 500) of random defect configurations for the finite 7-atom cross section wire and calculate the conductance at Fermi level. For each configuration, the defects are created by randomly selecting the location of the vacancies. Normalized counts of conductance are shown in Fig.6. Three major peaks at 1 $G_0$, 2.6 $G_0$, and 4 $G_0$ are found, in agreement with the experimental results. We find that the locations of these three peaks are insensitive to the length of finite nanowire length and total number of defect configurations. These features can be understood by the above discussions on the effects of various defects on crystalline nanowire. The peak at 4 $G_0$ corresponds to the case of one or few singe-atom defects that disrupt only one conduction channel from the perfect wire. The dominant peak at 1 $G_0$ could be associated with the presence of several multi-atoms defects, which disrupt the quantum conductance so dramatically that only one conduction channel is left. Between the two extreme cases, a peak at $\sim$ 2.5 $G_0$ is found both experimentally and theoretically. In our calculations, four less pronounced peaks around 1.5 $G_0$ and 3.4 $G_0$ are also obtained, which were not observed experimentally \cite{31}. These peaks may correspond to the energetically less-favorate configurations. 

In summary, three conduction channels are obtained for the crystalline silver nanowire with four atoms per cross section, corresponding to three $s$ bands crossing the Fermi level. One conductance channel is easily disrupted by an individual single-atom defect (vacancy or adatom), independent of the defect configurations. With two separated single-atom defects, quantum interference leads to oscillation of conductance versus the defect distance. In the presence of multiple-atoms defects, one conduction channel (at Fermi energy) remains robust. The calculated histogram of conductance for a finite silver nanowires (seven atoms per cross section, 2 nm long) with random defect configurations compares well with experiment. Our results show that the ultrathin silver wires, even with some structural defects, are excellent candidates in nanoelectronics.

This work is supported by the University Research Council of University of North Carolina at Chapel Hill, Office of Naval Research Grant No. N00014-98-1-0597 and NASA Ames Research Center. We acknowledge computational support from the North Carolina Supercomputer Center.
\ \\
\ \\
$^*$ E-mail: jpl@physics.unc.edu

\end{multicols}
\end{document}